\input harvmac 
\input pictex
\overfullrule=0pt

\def\sgn{\mathop{{\rm sgn} }}
\def\neqno#1{\eqnn{#1} \eqno #1}

\Title{hep-th/0006158, UTPT-00-07} {\vbox{
\centerline{Linearized Gravity about a Brane} }}
\centerline{\bf Hael Collins\footnote{$^\dagger$}{{\tt 
hael@physics.utoronto.ca}} and Bob 
Holdom\footnote{$^\ddagger$}{{\tt bob.holdom@utoronto.ca}} }
\medskip
\vbox{\it \centerline{Department of Physics}\vskip-2pt
\centerline{University of Toronto}\vskip-2pt
\centerline{Toronto, Ontario M5S 1A7, Canada}}

\bigskip
\bigskip
\centerline{Abstract}

\medskip {\baselineskip=10pt\centerline{\vbox{\hsize=.8\hsize\ninerm\noindent  
We use the Israel condition to treat carefully the weak-field perturbations
due to the presence of matter on a 3-brane embedded between two regions of
anti-de Sitter (AdS) space with different curvature lengths.  A four
dimensional Newton's Law only emerges at distances that are large compared to
the AdS lengths.  When a scalar curvature is included in the brane action,
however, it is possible to generate a four dimensional theory of gravity even
when one or both of the AdS lengths is large compared to distances along the
brane.  In particular, we provide an example in which the AdS lengths can be
larger than the millimeter experimental bound. }}}

\Date{June, 2000}
\baselineskip=12pt

\newsec{Introduction.}

One of the more intriguing new ideas in the past year has been the suggestion
that the physical universe could be embedded as a hypersurface in a higher
dimensional bulk space-time where the extra dimensions are non-compact.  The
fact that the Standard Model has been confirmed to scales approaching the
electroweak scale is accounted for in these models by assuming that these
fields are confined to this hypersurface, or `$3$-brane'.  Yet since gravity
describes the dynamics of space-time itself, some additional feature is needed
to produce an effectively four-dimensional theory of gravity, at least on
lengths greater than the millimeter scale probed.  In the scenario proposed by
Randall and Sundrum
\ref\rs{L.~Randall and R.~Sundrum, ``An alternative to compactification,''
Phys.\ Rev.\ Lett.\  {\bf 83}, 4690 (1999) [hep-th/9906064].},  
the bulk is five dimensional anti-de Sitter (AdS$_5$) space-time whose
curvature length provides the threshold above which the effective theory of
gravity is approximately Einstein's theory in four dimensions.  At distances
shorter than the AdS curvature length, the extra dimension is revealed.  

In studying extensions of the original Randall-Sundrum proposal, it is
important to have a method for describing the effective theory of gravity seen
by an inhabitant of the $3$-brane.  A natural approach 
\ref\gt{J.~Garriga and T.~Tanaka, ``Gravity in the brane-world,''
hep-th/9911055.}  
is to study the perturbations to the background metric produced when matter is
placed on the brane.  The problem is not as straightforward as it might seem
since the presence of brane-matter can alter the position of the brane in some
bulk coordinate systems.  In 
\ref\gkl{S.~B.~Giddings, E.~Katz and L.~Randall, ``Linearized gravity in brane
backgrounds,'' JHEP {\bf 0003}, 023 (2000) [hep-th/0002091].}  
it was shown that this `brane-bending' must be included to cancel the
appearance of a spurious scalar gravity term in the effective Einstein
equation in the original Randall-Sundrum scenario.  A better method was
recently introduced by 
\ref\gr{I.~Y.~Aref'eva, M.~G.~Ivanov, W.~Muck, K.~S.~Viswanathan and
I.~V.~Volovich, ``Consistent linearized gravity in brane backgrounds,''
hep-th/0004114.},  
building upon their earlier work 
\ref\oldiv{M.~G.~Ivanov and I.~V.~Volovich, ``Metric fluctuations in brane
worlds,'' hep-th/9912242 and W.~Muck, K.~S.~Viswanathan and I.~V.~Volovich,
``Geodesics and Newton's law in brane backgrounds,'' hep-th/0002132.},  
in which the gauge is chosen so that the brane remains flat, even when matter
is placed on the brane.  As it is presented in $\gr$, this method works best
when the model has an orbifold symmetry.  In this article, we shall extend the
formalism to permit the study of linearized gravity in asymmetric bulk
space-time backgrounds.

Central to our discussion is a careful treatment of the boundary conditions at
the brane.  In the previous analyses of linearized gravity in a brane
background, the energy-momentum contribution from the brane to the full bulk
Einstein equation was incorporated using a $\delta$-function source term. 
However, some of the functions that multiply this $\delta$-function become
ill-defined when the bulk space-time no longer has the standard orbifold
symmetry.  Fortunately, the approach introduced by Israel 
\ref\israelref{W.~Israel, ``Singular Hypersurfaces And Thin Shells In General
Relativity,'' Nuovo Cim.\  {\bf B44S10}, 1 (1966).}  
relates the discontinuities of functions evaluated on opposite sides of the
brane to functions that are well-defined on the brane.

We have also considered the effect of adding a scalar curvature term to the
brane action.  For our purpose, we shall simply regard this curvature term as
the next natural term in an effective theory expansion in powers of
derivatives.  Such a term can arise in some quantum field theories 
\ref\capper{D.~M.~Capper, ``On Quantum Corrections To The Graviton
Propagator,'' Nuovo Cim.\  {\bf A25}, 29 (1975).}  
\ref\adler{S.~L.~Adler, 
Phys.\ Rev.\ Lett.\  {\bf 44}, 1567 (1980); 
S.~L.~Adler, 
Phys.\ Lett.\  {\bf B95}, 241 (1980); 
S.~L.~Adler, 
Rev.\ Mod.\ Phys.\  {\bf 54}, 729 (1982); Erratum-ibid. {\bf 55}, 837 (1983).} 
\ref\zee{A.~Zee, ``Calculating Newton's Gravitational Constant In Infrared
Stable Yang-Mills Theories,'' Phys.\ Rev.\ Lett.\  {\bf 48}, 295 (1982).}   
from radiative corrections to gravity that involve fields confined to the
brane circulating in the loops 
\ref\dgp{G.~Dvali, G.~Gabadadze and M.~Porrati, ``4D gravity on a brane in 5D
Minkowski space,'' hep-th/0005016.}.  
Scalar curvature terms also occur extensively in the study of the AdS/CFT
correspondence where they are used to regularize the bulk action 
\ref\adscft{See for example R.~Emparan, C.~V.~Johnson and R.~C.~Myers,
Phys.\ Rev.\  {\bf D60}, 104001 (1999) [hep-th/9903238] and V.~Balasubramanian
and P.~Kraus, 
Commun.\ Math.\ Phys.\  {\bf 208}, 413 (1999) [hep-th/9902121].}.   
Moreover, the brane curvature term is required for the emergence of $4d$
gravity in theories in which the brane is embedded in a flat bulk space-time
$\dgp$.  

The presence of a curvature term in the brane action leads to a particularly
intriguing possibility.  Since an effectively $4d$ theory of gravity from the
bulk arises at distances along the brane greater than the AdS lengths, by
adding a curvature term on the brane which has the dominant contribution to
gravity at or below these scales, we can obtain a standard $4d$ Einstein
theory of gravity valid at all testable scales.  With these ingredients, we
can find a realistic scenario in which the AdS lengths can be above the
millimeter scale and the bulk Planck scale is in the TeV range at the cost of
mildly fine-tuning the coefficient of the brane curvature term.

The next section contains a detailed derivation of the weak-field perturbation
to the background metric due to fields placed on the brane.  Although some of
this material builds on the elegant formalism of $\gr$, we have included the
derivation of the bulk equations along with the new material describing the
behavior at the boundary for completeness.  After a short aside on the gauge
invariance of our results, in section 4 we use our results to study the
effective theory of gravity along the brane for a universe consisting of a
$3$-brane separating two regions of AdS$_5$ with different curvatures.  In
section 5 we introduce a scalar curvature term into the action and show that
it can provide a source for four dimensional gravity even in theories that do
not have a $4d$ Newton's Law without it.  Section 6 concludes with some
comments on size of the $5d$ Planck mass.

\newsec{Linearized Gravity in an Asymmetric Background.}

The action for a $3$-brane embedded in a five dimensional bulk space-time can
be divided into bulk and brane components.  For the bulk, we use an
Einstein-Hilbert action while on the brane we initially consider a minimal
action containing a surface tension term plus a term for fields confined to
the brane:
$$\eqalign{
S &= M_5^3\int d^4xdy\, \sqrt{-g}\, \left( 2\Lambda_{\sgn(y)} + R \right) 
+ 2M_5^3\int_{\rm brane} d^4x\, \sqrt{-h}\, \Delta K \cr
&\qquad + M_5^3\int_{\rm brane} d^4x\, \sqrt{-h}\, \left( - {12\over\ell} +
{1\over M_5^3}\, {\cal L}_{\rm fields} \right) . \cr} \neqno\action$$
$g_{ab}$ is the metric for the bulk space-time and $h_{ab}$ is the induced
metric along the $3$-brane.  $M_5$ is the bulk Planck mass.  We shall denote
the coordinates along the brane by $x^\mu$, where $\mu,\nu,\cdots = 0,1,2,3$;
the fifth coordinate $y$ is chosen so that the brane lies at $y=0$ and
$a,b,c,\ldots=0,1,2,3,y$.  Notice that we have allowed the cosmological
constant to have different values to either side of the brane, represented by
the $\sgn(y)$-dependence, which leads to different metrics in the two bulk
regions.  When the metric changes between regions of the bulk, it is necessary
to include a term in the boundary action that depends on the trace of the
extrinsic curvature, $K=h^{ab}K_{ab}$, defined below.  Since the boundary for
the two bulk regions is provided by the brane itself, we include a term in the
brane action for the difference in the extrinsic curvature for the two
regions, $\Delta K_{ab} \equiv K_{ab}|_{y=0^-} - K_{ab}|_{y=0+}$.  For
comparison, in the standard Randall-Sundrum universe $\rs$ $\Lambda=6/\ell^2$
when the bulk metric is $ds^2 = e^{-2|y|/\ell}\eta_{\mu\nu}\, dx^\mu dx^\nu +
dy^2$ for ${\cal L}_{\rm fields}=0$.  $\eta_{\mu\nu}=\hbox{diag}\, (-1,1,1,1)$
is the four dimensional Minkowski metric.

The approach that we shall use to extract the theory of gravity for an
observer on the $3$-brane is to study gravity in the weak field limit,
expanding in a small perturbation about the classical background.  The
treatment of linearized gravity was first presented in $\gt$, and then more
fully in $\gkl$.  In both these papers the gauge was chosen so that the bulk
metric assumed the form 
$$g_{ab} = \pmatrix{ e^{-2|y|/\ell} \eta_{\mu\nu} + \gamma_{\mu\nu}  &0\cr  0
&1\cr}
\neqno\stdgauge$$
where $\gamma_{\mu\nu}$ is regarded as a small perturbation.  The remaining
gauge freedom is used to choose $\partial^\mu \gamma_{\mu\nu}=0$.  In these
coordinates the presence of matter on the brane distorts its position in the
bulk space so that it no longer lies at $y=0$.  In geometries without the
usual orbifold symmetry, this ``brane bending'' leads to subtleties in
correctly imposing boundary conditions at the position of the brane.  As a
consequence, it is convenient to relax these gauge conditions  and instead to
work in a gauge where the position of the brane remains fixed at $y=0$.

The problem of finding such a gauge in order to study linearized gravity was
addressed in $\gr$.  There, the metric tensor is written using the
time-slicing formalism
\ref\mtw{C.~W.~Misner, K.~S.~Thorne and J.~A.~Wheeler, {\it Gravitation\/},
(W. H. Freeman:  San Francisco) 1973.}  
as
$$g_{ab} = \pmatrix{ \hat g_{\mu\nu}  &n_\mu\cr  n_\nu &n^2+n_\lambda
n^\lambda\cr}
\qquad\qquad
g^{ab} = {1\over n^2} \pmatrix{ \hat g^{\mu\nu}  &-\hat
g^{\mu\lambda}n_\lambda\cr  -\hat g^{\nu\lambda}n_\lambda &1\cr}
\neqno\tsgauge$$
where 
$$\hat g_{\mu\nu} = f(y) (\eta_{\mu\nu} + \gamma_{\mu\nu}) . \neqno\fdef$$  
$n$ and $n_\mu$ are respectively called the lapse function and the shift
vector.  Here we have written a more general warp factor, $f(y)$, to allow
configurations in which the brane separates two regions of AdS$_5$ with
different cosmological constants.  In the limit in which the perturbations
about the background are small, we consider $\gamma_{\mu\nu}$, $n_\mu$ and
$\phi\equiv n^2-1$ all to be small quantities of roughly the same size.  

The advantage of this metric is that it admits a foliation of the bulk space
by hypersurfaces where $y$ is constant.  The normal vector to any of these
hypersurfaces can be written as 
$$N_a = (0,0,0,0,-n) \qquad N^a = {1\over n} (\hat g^{\mu\nu} n_\nu, -1) 
\neqno\normals $$
and the induced metric along the surface is given by 
$$h_{ab} = g_{ab} - N_aN_b = \pmatrix{ \hat g_{\mu\nu} & n_\mu\cr  n_\nu
&n_\lambda n^\lambda\cr}
\qquad
h^{ab}  = {1\over n^2} \pmatrix{ \hat g^{\mu\nu}  - \hat g^{\mu\lambda} \hat
g^{\nu\rho} n_\lambda n_\rho &0\cr  0 &0\cr}
\neqno\imetric$$
In addition to the curvature induced by this metric along a hypersurface, the
bulk space also produces an extrinsic curvature structure defined by the
symmetric tensor\foot{Note that we have chosen our sign to agree with that
used in $\gr$ which is the  opposite of that used in 
\ref\ch{H.~Collins and B.~Holdom, ``Brane cosmologies without orbifolds,''
hep-ph/0003173.}  .}  
$$K_{ab} = - h_a^{\ c}\nabla_c N_b \neqno\extrinsicdef$$
where the covariant derivative is with respect to the bulk metric.  To leading
order in the perturbations, the extrinsic curvature is then 
$$K_{\mu\nu} = {1\over 2n} \left[ \partial_y\hat g_{\mu\nu} - \partial_\mu
n_\nu - \partial_\nu n_\mu \right] + \cdots, \neqno\extrtrans$$
with $K_{\mu y}= {1\over 2}{f'\over f}n_\mu$ and $K_{yy}=0$.

One of the new features in a bulk space-time without an orbifold symmetry is
that some quantities become ill-defined on the brane.  For example, when the
cosmological constant differs on the opposite sides of the brane, it changes
discontinuously at $y=0$.  One solution to this problem is to allow the brane
to have some finite thickness, so that the transition from one region to
another proceeds smoothly.  Alternatively, we can use the Israel junction
equations which relate changes in bulk quantities to the brane tension and the
energy-momentum tensor for fields confined to the brane.  This approach avoids
evaluating potentially ill-defined functions on the brane.  

The behavior of gravity in the bulk is governed by the usual Einstein
equation,
$$R_{ab} - {1\over 2} R g_{ab} = g_{ab} \Lambda_{\sgn(y)} . \neqno\bulkE$$
At the brane, the change in the extrinsic curvature is determined by the
Israel condition $\ch$,
$$\Delta K_{ab} = 8\pi \left[ T_{ab} - {1\over 3} h_{ab} T \right] .
\neqno\israelall$$
$T_{ab}$ includes both the brane tension as well as fields living on the
brane.  Although we shall use the Israel condition to fix the behavior of the
metric perturbations at the brane, it is nevertheless useful to write $\bulkE$
in the Gauss-Codacci form by decomposing it into the transverse and orthogonal
components with respect to an arbitrary $y=\hbox{constant}$ hypersurface,
$$\eqalign{
\hat R + K^\mu_\nu K^\nu_\mu - K^2 &= 2\Lambda_{\sgn(y)} \cr
\partial_\mu K - \hat\nabla_\nu K^\nu_\mu &= 0 \cr
R_{\mu\nu} &= {2\over 3} \hat g_{\mu\nu} \Lambda_{\sgn(y)} . \cr}
\neqno\codacci$$
Here $\hat R$ is the scalar curvature associated with $\hat g_{\mu\nu}$ and is
calculated in the appendix.  Away from the brane, $y\not = 0$, the zeroth
order terms from $\codacci$ determine the warp factor which appears in the
unperturbed metric,\foot{Since we are writing only the equations for the bulk,
we have not included the usual $\delta$-function source term for the brane in
the $f^{\prime\prime}$ equation.}
$${f^{\prime 2}\over f^2} = - {2\over 3} \Lambda_{\sgn(y)} \qquad\qquad
{f^{\prime\prime}\over f} = - {2\over 3} \Lambda_{\sgn(y)} . \neqno\einzero$$
In this more general picture, $\Lambda_{\sgn(y)}$ can have different values
for $y>0$ and $y<0$ so that $f^{\prime 2}$ becomes ill-defined on the brane.  

The terms linear in the perturbations for the first two Gauss-Codacci
equations $\codacci$ impose constraints on the metric:
$$\eqalign{
{1\over f} \left[ \partial^\mu \partial^\nu \gamma_{\mu\nu} -
\partial^\lambda\partial_\lambda \gamma + 3{f'\over f} \partial^\lambda
n_\lambda \right] &= - 3\phi {f^{\prime 2}\over f^2} + {3\over 2} {f'\over f}
\partial_y\gamma \cr
{1\over f} \left[ \partial^\lambda\partial_\lambda n_\mu -
\partial_\mu\partial^\nu n_\nu \right] &= {3\over 2} {f'\over f}
\partial_\mu\phi + \partial_y \left[ \partial^\nu\gamma_{\mu\nu} -
\partial_\mu \gamma \right] . \cr}
\neqno\gcact$$
The important feature of these expressions is the appearance of the $f^{-1}$
factors on their left sides.  For regions of anti-de Sitter space, $f^{-1}$
typically grows exponentially as we move farther into the bulk so that at some
distance away from the brane, the assumption that $\gamma_{\mu\nu}$ and $\phi$
are small breaks down, unless we use gauge freedom to set the right sides of
both expressions to zero.  Therefore, in the bulk we choose, similarly to
$\gr$,
$$\phi = {1\over 2}{f\over f'}\partial_\mu\gamma 
\qquad\qquad
\partial^\mu\tilde\gamma_{\mu\nu} = 0 \neqno\gaugechoice$$
where as $\tilde\gamma_{\mu\nu}$ denotes the traceless part of
$\gamma_{\mu\nu}$:
$$\tilde\gamma_{\mu\nu} \equiv \gamma_{\mu\nu} - {1\over 4}
\eta_{\mu\nu}\gamma . \neqno\tracelessdef$$
For these equations, $\gcact$ constrains the shift vector to be of the
following form:  
$$n_\mu = {1\over 4} {f\over f'}\partial_\mu\gamma + A_\mu .
\neqno\fixshift$$
Here $A_\mu$ is a free vector field ($\lform A_\mu = 0$) with vanishing
$4$-divergence ($\partial^\mu A_\mu =0$).  Later we shall use the gauge
freedom to choose $A_\mu=0$.  

For simplicity, we assume that all the matter is confined to the brane.  On
the brane, the energy-momentum tensor receives contributions from both the
brane tension, which we write as $-{6\over\ell}$, as well as a term
$t_{\mu\nu}$ for the fields confined to the $3$-brane,
$$T_{\mu\nu} = -{3\over 4\pi} {1\over\ell} \hat g_{\mu\nu} + t_{\mu\nu}(x) .
\neqno\SEbrane$$
The energy-momentum $t_{\mu\nu}$ determines the behavior of $\gamma_{\mu\nu}$
so it should also be treated as a small quantity.  The surface tension of the
brane is related to the discontinuity in the slope of the warp factor at the
brane.  Explicitly, the unperturbed piece of the Israel condition requires the
usual fine-tuning of the brane tension,
$$f'|_{y=0-} - f'|_{y=0+} \equiv \Delta [f'] = {4\over\ell} f .
\neqno\finetune$$

The behavior of $\gamma_{\mu\nu}$ in the bulk is found using the transverse
components of the Einstein equation $\codacci$,
$$\eqalign{
&\hat R_{\mu\nu} + {1\over 2} {f'\over f} \left[ \partial_\mu n_\nu +
\partial_\nu n_\mu + \eta_{\mu\nu} \partial^\lambda n_\lambda \right] -
{1\over 2} f' \left[ 2\partial_y \gamma_{\mu\nu} + {1\over 2} \eta_{\mu\nu}
\partial_y\gamma - {1\over 2} \eta_{\mu\nu} \partial_y\phi \right] \cr
& - {1\over 2} \partial_\mu\partial_\nu\phi + {1\over 2} \left[
f^{\prime\prime} + {f^{\prime 2}\over f} \right] \eta_{\mu\nu}\phi - {1\over
2} f \partial_y^2\gamma_{\mu\nu} + {1\over 2} \partial_y \left[ \partial_\mu
n_\nu + \partial_\nu n_\mu \right] = 0 , \cr}
\neqno\gcacttrans$$
where we have substituted the equations determining the warp factor
$\einzero$.  The transverse components of the Israel condition specify the
boundary conditions at the brane,
$$\Delta \left[ - {1\over 2} f' \phi\eta_{\mu\nu} + f
\partial_y\gamma_{\mu\nu} - \partial_\mu n_\nu - \partial_\nu n_\mu \right] =
16\pi \left[ t_{\mu\nu} - {1\over 3} \eta_{\mu\nu} t \right] 
\neqno\israelfirst$$
where we have imposed the zeroth order constraint $\finetune$.  
In both these expressions we have retained only the linear terms in the metric
perturbation.  Both $\phi$ and $n_\mu$ can be eliminated from these equations
through $\gaugechoice$ and $\fixshift$ so that
$$\partial_y (f\partial_y\tilde\gamma_{\mu\nu}) + f'\partial_y
\tilde\gamma_{\mu\nu} + \lform \tilde\gamma_{\mu\nu} 
- {f'\over f} \left[ \partial_\mu A_\nu + \partial_\nu A_\mu \right] -
\partial_y \left[ \partial_\mu A_\nu + \partial_\nu A_\mu \right] = 0 
\neqno\gcactb$$
for $y\not = 0$ and
$$\Delta \left[ f \partial_y\tilde\gamma_{\mu\nu} 
- {1\over 2} {f\over f'} \partial_\mu\partial_\nu\gamma 
- \partial_\mu A_\nu - \partial_\nu A_\mu \right] 
= 16\pi \left[ t_{\mu\nu} - {1\over 3} \eta_{\mu\nu} t \right]
\neqno\israela$$
at the brane, $y=0$.
Finally by making a gauge transformation to eliminate $A_\mu$ and noting that
since the metric must be continuous at the brane, so that both $f$ and
$\gamma_{\mu\nu}$ are continuous there, we have the following equations for
the behavior of $\gamma_{\mu\nu}$,
$$\partial_y (f\partial_y\tilde\gamma_{\mu\nu}) + f'\partial_y
\tilde\gamma_{\mu\nu} + \lform \tilde\gamma_{\mu\nu}  = 0 \neqno\gammabulk$$
in the bulk and
$$f \Delta \left[ \partial_y\tilde\gamma_{\mu\nu} \right] 
- {1\over 2} \Delta \left[ {f\over f'} \right] \partial_\mu\partial_\nu\gamma 
= 16\pi \left[ t_{\mu\nu} - {1\over 3} \eta_{\mu\nu} t \right]
\neqno\gammabrane$$
on the brane.

As in $\gr$, we can eliminate the $\gamma$ from $\gammabrane$ by taking the
trace of this expression
$$\Delta \left[ {f\over f'} \right] \lform\gamma = {32\pi\over 3} t
\neqno\gammatrace$$
and inverting the operator $\lform$ to write formally 
$$f\partial_y\tilde\gamma_{\mu\nu}|_{y=0^+} -
f\partial_y\tilde\gamma_{\mu\nu}|_{y=0^-} = - 16\pi \left[ t_{\mu\nu} -
{1\over 3} \left( \eta_{\mu\nu} - {\partial_\mu\partial_\nu\over\lform}
\right) t \right] . \neqno\gammabraneexp$$
As for the case of the orbifold, taking the $4$-divergence of this expression
shows that the energy-momentum tensor for the brane fields is conserved,
$\partial^\mu t_{\mu\nu}=0$.

\newsec{Gauge Invariance.}

Before considering the weak field behavior for some specific geometries, we
should show that both the bulk equation for $\tilde\gamma_{\mu\nu}$ $\gcactb$
as well as the boundary condition $\israela$ are left invariant under changes
of coordinates that still preserve the position of the brane.  Under a small
change of coordinates, 
$$x^{\prime\mu} = x^\mu - \xi^\mu(x,y) \qquad\qquad y' = y - \xi^5(x,y); 
\neqno\smallchange$$
the statement that the position of the brane is left invariant is that
$\xi^5(x,0)=0$.  The components of the metric change according to 
$$\eqalign{
\gamma_{\mu\nu} &= \gamma'_{\mu\nu} - {f'\over f} \eta_{\mu\nu} \xi^5 -
\partial_\mu\xi_\nu - \partial_\nu\xi_\mu \cr
n_\mu &= n'_\mu - f \partial_y\xi_\mu - \partial_\mu\xi^5 \cr
\phi &= \phi' - 2 \partial_y\xi^5 .\cr}
\neqno\gaugerules$$
The trace and the traceless part of $\gamma_{\mu\nu}$ transform as 
$$\eqalign{
\gamma &= \gamma' - 4 {f'\over f}\xi^5 - 2 \partial^\mu\xi_\mu \cr
\tilde\gamma_{\mu\nu} &= \tilde\gamma'_{\mu\nu} - \partial_\mu\xi_\nu -
\partial_\nu\xi_\mu + {1\over 2}\eta_{\mu\nu}\partial^\lambda\xi_\lambda .\cr}
\neqno\gaugemetric$$
If we maintain the same gauge conditions $\gaugechoice$ before and after the
coordinate transformation, we have further that
$$\eqalign{
\partial_y\partial^\lambda\xi_\lambda &= - 2 \xi^5\partial_y\left( {f'\over f}
\right) \cr 
\lform\xi_\mu + {1\over 2} \partial_\mu \partial^\nu\xi_\nu &= 0 . \cr}
\neqno\atransform$$
Lastly, we note that the vector field $A_\mu$ transforms as follows:
$$A_\mu = A'_\mu - f \partial_y\xi_\mu + {1\over 2} {f\over f'}\partial_\mu
\partial^\nu\xi_\nu . \neqno\vectortransform$$
We have used this residual gauge freedom to set $A_\mu=0$.  

The bulk equation $\gcactb$ is invariant provided that 
$$\left[ \partial_\mu\partial_\nu\partial_\lambda \xi^\lambda + 2f'\xi^5
\eta_{\mu\nu} - 2 {f\over f'}\partial_\mu\partial_\nu\xi^5 \right] \partial_y
\left( {f'\over f}\right) + f\eta_{\mu\nu} \partial_y \left[ \xi^5 \partial_y
\left( {f'\over f} \right) \right] = 0. \neqno\invariantbulk$$
$\einzero$ implies that $f'/f=\hbox{constant}$ for $y\not=0$, so
$\invariantbulk$ is satisfied.

Under a coordinate transformation, we also find that the left side of the
Israel condition $\israela$ is invariant when 
$$\Delta \left[ \partial_\mu\partial_\nu\xi_5 \right] - 2 f\eta_{\mu\nu}
\Delta \left[ \xi^5\partial_y\left( {f'\over f}\right)  \right] = 0 .
\neqno\gtisrael$$
These quantities are evaluated in the bulk on either side of the brane so as
we have seen the second term vanishes.  $\partial_\mu\partial_\nu\xi_5$ also
vanishes since $\xi^5(x,y)\to 0$ as $y\to 0$, under our constraint that the
position of the brane remains fixed.  The gauge invariance of the Israel
condition demonstrates that $\gamma(x,0)$ is a gauge-invariant quantity
determined only by the physical fields placed on the brane through
$\gammatrace$.  This result is important since, as we shall see in the next
section, $\gamma(x,0)$ enters the expression for the effective Einstein
equations on the brane.

Before concluding this aside on the gauge invariance of $\gcactb$--$\israela$,
we should mention how the fixed-brane gauge is related to the bent-brane
gauges of $\gt$ and $\gkl$.  The appearance of the term $\Delta \left[
\partial_\mu\partial_\nu\xi_5 \right]$ in the gauge transformed Israel
condition shows that if the position of the brane is not kept fixed, then
gauge-dependent terms will appear in the expression for $\gamma$.  Since these
terms need to be subtracted from the final expression for the effective
theory, it is convenient to work in a gauge in which they are absent.  More
importantly for a universe without an orbifold symmetry, the bending upon
either side of the brane might not be equal so that extra care must be made to
properly describe the boundary conditions at the brane.\foot{This point is
stressed in the note accompanying equation $(3.11)$ of $\gkl$.}  Again, this
complication is avoided when the brane's position is fixed.

\newsec{An Example:  A Brane between Two Regions of AdS$_5$.}

As an application, we examine in depth the behavior of a universe in which a
brane with tension $-6/\ell$ separates two regions of AdS$_5$ with different
curvatures, 
$$\Lambda = \cases{ 6/\ell_1^2 &for $y>0$\cr 6/\ell_2^2 &for $y<0$\cr},
\neqno\ccdef$$
for which the corresponding warp factor is $f(y)=e^{-2y/\ell_1}$ for $y>0$ and 
$f(y)=e^{2y/\ell_2}$ for $y<0$.  The fine-tuning condition for the brane
tension $\finetune$ is 
$${1\over\ell} = {1\over 2} \left[ {1\over\ell_1} + {1\over\ell_2} \right] .
\neqno\finetuneeg$$
Since $4$-derivatives appear in the equation determining
$\tilde\gamma_{\mu\nu}$, it is convenient to Fourier transform in the
directions along the brane so that in the region $y>0$ $\tilde\gamma_{\mu\nu}$
satisfies
$$\partial_y (e^{-2y/\ell_1}\partial_y\tilde\gamma_{\mu\nu}) - {2\over\ell_1}
e^{-2y/\ell_1} \partial_y \tilde\gamma_{\mu\nu} - p^2 \tilde\gamma_{\mu\nu} =
0 \neqno\rightregion$$
with an analogous expression for $y<0$ involving $\ell_2$.  For $p^2>0$, the
solution to this differential equation is in terms of Bessel functions,
$$\tilde\gamma_{\mu\nu}(p,y>0) = c^+_{\mu\nu}(p)\, e^{2y/\ell_1} \left[ 
K_2(e^{y/\ell_1}\ell_1|p|) + A\, I_2(e^{y/\ell_1}\ell_1|p|) \right] .
\neqno\rightgensoln$$
Since $I_2(e^{y/\ell_1}|p|\ell_1)$ diverges as $y\to\infty$, we set $A=0$. 
From $\gammabraneexp$ and imposing continuity of the metric across the brane,
we discover the following behavior for weak fluctuations about the background
solution 
$$\eqalign{
\tilde\gamma_{\mu\nu}(p,y) = {16\pi\over |p|} &\left[ t_{\mu\nu} - {1\over 3}
\left[ \eta_{\mu\nu} - {p_\mu p_\nu\over p^2} \right] t \right] \cr
&\times
{e^{2y/\ell_1} K_2(\ell_2|p|) K_2(\ell_1|p| e^{y/\ell_1}) \over 
K_1(\ell_1|p|) K_2(\ell_2|p|) + K_1(\ell_2|p|) K_2(\ell_1|p|) }
\cr}\neqno\rightsoln$$
for $y>0$ and
$$\eqalign{
\tilde\gamma_{\mu\nu}(p,y) = {16\pi\over |p|} & \left[ t_{\mu\nu} - {1\over 3}
\left[ \eta_{\mu\nu} - {p_\mu p_\nu\over p^2} \right] t \right] \cr
&\times 
{e^{-2y/\ell_2} K_2(\ell_1|p|) K_2(\ell_2|p| e^{-y/\ell_2}) \over 
K_1(\ell_1|p|) K_2(\ell_2|p|) + K_1(\ell_2|p|) K_2(\ell_1|p|) }
\cr}\neqno\leftsoln$$
for $y<0$.  The value of the trace $\gamma$ at the brane is determined by the
Israel condition $\gammatrace$ to be 
$$\gamma(p,0) = -{32\pi\over 3 p^2} {2\over\ell_1+\ell_2}\, t ,
\neqno\gammatraceeg$$
and along with $\tilde\gamma_{\mu\nu}$ is needed to find the effective theory
of gravity seen by an observer confined to the brane.

A possible source for a discrepancy between a purely four dimensional theory
of gravity and the effective low energy theory of gravity for an inhabitant of
the $3$-brane is in the appearance of deviations from the Einstein equation. 
The intrinsic Einstein tensor is given by $\gr$
$$\hat R_{\mu\nu} - {1\over 2}\eta_{\mu\nu} \hat R = - {1\over 2} \lform
\tilde\gamma_{\mu\nu} - {1\over 4} \left( \partial_\mu\partial_\nu\gamma -
\eta_{\mu\nu}\lform\gamma \right) . \neqno\fourdEIN$$
As a first case, we look in the limit in which both of the lengths associated
with the bulk AdS$_5$ regions are small compared to the typical length scale
probed on the brane, $\ell_1|p|,\ell_2|p|\ll 1$.  Using $K_2(z)=2
z^{-2}+\cdots$ and $K_1(z)=z^{-1}+\cdots$ for small $z$, we have
$$\tilde\gamma_{\mu\nu}(p,0) = {16\pi\over p^2} \left[ t_{\mu\nu} - {1\over 3}
\left[ \eta_{\mu\nu} - {p_\mu p_\nu\over p^2} \right] t \right] 
{2\over \ell_1+\ell_2 } + \cdots ; \neqno\zeromode$$
combining this with the trace $\gammatraceeg$, we recover the standard theory: 
$$\hat R_{\mu\nu} - {1\over 2}\eta_{\mu\nu} \hat R = {16\pi\over\ell_1+\ell_2}
t_{\mu\nu} + \cdots . \neqno\fourdEINsmall$$
Later we shall discuss theories that deviate from this standard behavior.

We also examine whether a stationary point mass $M$ on the brane produces a
$1/r$ Newtonian potential $\gr$.  For a point source at rest on the brane, the
energy-momentum tensor has only one non-vanishing component: 
$$t_{00}(x) = {M\over M_5^3} \delta^3(\vec x) \qquad\hbox{or}\qquad
t_{00}(p) = 2\pi {M\over M_5^3} \delta(p_0) \neqno\pointsource$$
where $M_5$ is the five dimensional Planck mass.  When the velocity of a test
mass is small, the gravitational field is stationary and weakly perturbed away
from the background metric, we can extract the Newtonian potential by
following the motion of a test mass along a geodesic.  The leading piece of
the geodesic equation in this Newtonian limit is 
$${d^2x^i\over dt^2} = \hat\Gamma^i_{00} = - {1\over 2}\partial_i\gamma_{00} =
- {1\over 2}\partial_i\left( \tilde\gamma_{00} - {1\over 4}\gamma \right)
\neqno\newtonlimit$$
from which we can extract the ordinary Newtonian potential,
$${d^2x^i\over dt^2} = \partial_i V .  \neqno\newtonpot$$
Typically in brane-world scenarios the standard model fields are confined to
the brane by some unspecified mechanism so that a test mass is not free to
move along an arbitrary geodesic, but one constrained to lie within the brane. 
Thus in the geodesic equation, we have used the Christoffel symbols for the
{\it induced\/} metric.  

Returning to the limit where $\ell_1|p|,\ell_2|p|\ll 1$, we note that the
presence of a point mass on the brane produces a perturbation of 
$$\tilde\gamma_{00}(p,0) \approx {32\pi\over 3 p^2} {2\over \ell_1+\ell_2 }
t_{00}  \neqno\zeromodett$$
while the trace $\gamma$ is fixed by $\gammatraceeg$.  Thus when both the
curvatures of the AdS regions are small compared to typical lengths probed on
the brane, we recover a standard attractive $1/r$ potential,\foot{If the test
mass were allowed to move into the bulk we would have used $\Gamma^i_{00}$
rather than $\hat\Gamma^i_{00}$ in $\newtonlimit$, which has the effect of
removing the trace term, and the Newtonian potential would have been enhanced
by a factor of $4/3$.}
$$V(r) = - {1\over 2} \int {d^4p\over (2\pi)^4}\, e^{-ip\cdot x} \gamma_{00} =
- {M\over M^3_5} {2\over \ell_1+\ell_2 } {1\over r} .  
\neqno\potsmall$$

The preceding example represents only a slight modification of the standard
Randall-Sundrum scenario which corresponds to setting $\ell_1=\ell_2=\ell$. 
However, we can also study alternative limits such as a universe in which one
of the AdS lengths is small compared to the typical scales probed on the
brane, $\ell_1|p|\ll 1$, but where the other is much larger, $\ell_2|p|\gg 1$. 
This case can be taken as an approximation to a half-AdS, half-flat universe. 
In this limit, the leading contribution to the traceless piece of the metric
perturbation is 
$$\tilde\gamma_{\mu\nu}(p,0) = {16\pi\over |p|} \left[ t_{\mu\nu} - {1\over 3}
\left[ \eta_{\mu\nu} - {p_\mu p_\nu\over p^2} \right] t \right] 
{1\over 1 + {1\over 2}\ell_1|p| - {3\over 2} (\ell_2|p|)^{-1}} + \cdots 
\neqno\intmode$$
where we have used 
$${K_1(\ell_2|p|)\over K_2(\ell_2|p|)} = 1 - {3\over 2} {1\over\ell_2|p|} +
{\cal O}\left( {1\over\ell_2^2p^2} \right) \neqno\bigK$$
and neglected higher order terms as $\ell_2|p|\to\infty$.  Since the leading
behavior is $|p|^{-1}$ and not $p^{-2}$, the leading contribution to the
Newtonian potential has a $1/r^2$ behavior:
$$V(r) = - {8\over 3\pi} {M\over M^3_5} {1\over r^2} 
- {4\over 3}{M\over M^3_5} {1\over\ell_2} {1\over r} 
+ {4\over 3\pi} {M\over M^3_5} {\ell_1^2\over r^4} + \cdots . \neqno\potint$$
In the limit $r\ll\ell_2$, the second term is actually a subleading
correction.  This example confirms the expectation that when the brane borders
a flat $5d$ region, it is revealed by a $1/r^2$ Newtonian potential.

\newsec{A Scalar Curvature Term in the Brane Action.}

We have used so far a minimal action on the brane consisting of only a surface
tension and a lagrangian for the fields confined to the brane.  Yet from an
effective field theory approach to the brane action, higher order terms could
be present, suppressed by powers of derivatives, that involve powers of the
scalar curvature associated with the induced metric on the brane.  We shall
focus on the simplest such term, a scalar curvature on the brane, whose
presence can produce a $4d$ Newton's Law even in theories in which the brane
borders a flat bulk region.

The presence of a scalar curvature term in the brane action, which we write
with an arbitrary dimensionless coefficient $b$ by extracting a factor of the
length scale associated with the brane tension, 
$$S_{\rm brane} = M_5^3 \int_{\rm brane} d^4x\, \sqrt{-h}\, \left( -
{12\over\ell} + {1\over 2} b\ell\, {\cal R} + {1\over M_5^3}\, {\cal L}_{\rm
fields} \right) , \neqno\baction$$
introduces a new term in the Israel condition $\ch$, 
$$\Delta K_{\mu\nu} = {2\over\ell} + 8\pi \left[ t_{\mu\nu} - {1\over 3}
\eta_{\mu\nu} t \right] - b {\ell\over 2} \left[ {\cal R}_{\mu\nu}  - {1\over
6} h_{\mu\nu} {\cal R} \right] .\neqno\israelR$$
${\cal R}_{ab}$ and ${\cal R}$ are the Ricci tensor and curvature scalar for
the induced metric $h_{ab}$.  To leading order in the perturbations, these
curvatures are given by ${\cal R}_{\mu\nu} = \hat R_{\mu\nu}$ and ${\cal R} =
\hat R$; their components are evaluated in the appendix.  The new boundary
condition imposed at $y=0$ is then 
$$f \Delta \left[ \partial_y\tilde\gamma_{\mu\nu} \right] 
- {1\over 2} \Delta \left[ {f\over f'} \right] \partial_\mu\partial_\nu\gamma 
= 16\pi \left[ t_{\mu\nu} - {1\over 3} \eta_{\mu\nu} t \right] 
+ {1\over 2} b\ell  \left[ \lform\tilde\gamma_{\mu\nu} 
+ {1\over 2} \partial_\mu\partial_\nu \gamma \right] .
\neqno\gammabraneR$$
As before, we can remove the $\gamma$ terms by taking the trace of
$\gammabraneR$, 
$$\left( \Delta \left[ {f\over f'} \right] + {1\over 2} b\ell \right)
\lform\gamma = {32\pi\over 3} t ,
\neqno\gammatraceR$$
which yields a condition on the traceless part of the metric perturbation
similar to that in $\gammabraneexp$ except that an additional term with the
$4$-Laplacian of $\tilde\gamma_{\mu\nu}$ appears,
$$f\partial_y\tilde\gamma_{\mu\nu}|_{y=0^+} -
f\partial_y\tilde\gamma_{\mu\nu}|_{y=0^-} = {1\over 2} b\ell \,
\lform\tilde\gamma_{\mu\nu} - 16\pi \left[ t_{\mu\nu} - {1\over 3} \left(
\eta_{\mu\nu} - {\partial_\mu\partial_\nu\over\lform} \right) t \right] .
\neqno\gammabraneexpR$$
For a universe with two regions of AdS$_5$ separated by a $3$-brane at $y=0$,
the value of the Fourier transform of the trace $\gamma$ at the brane is given
by 
$$\gamma = -{32\pi\over 3 p^2} {2\over\ell_1 + \ell_2 + b\ell}\,  t .
\neqno\gammatraceegR$$
The solution to the bulk equation $\gammabulk$ that satisfies this new
boundary condition has the form
$$\eqalign{
\tilde\gamma_{\mu\nu}(p,y) &= {16\pi\over |p|} \left[ t_{\mu\nu} - {1\over 3}
\left[ \eta_{\mu\nu} - {p_\mu p_\nu\over p^2} \right] t \right] \cr
&\times
{e^{2y/\ell_1} K_2(\ell_2|p|) K_2(\ell_1|p| e^{y/\ell_1}) \over 
K_1(\ell_1|p|) K_2(\ell_2|p|) + K_1(\ell_2|p|) K_2(\ell_1|p|) + {1\over
2}b\ell |p|\, K_2(\ell_1|p|) K_2(\ell_2|p|)} \cr}\neqno\rightsolnR$$
for $y>0$ and
$$\eqalign{
\tilde\gamma_{\mu\nu}(p,y) &= {16\pi\over |p|} \left[ t_{\mu\nu} - {1\over 3}
\left[ \eta_{\mu\nu} - {p_\mu p_\nu\over p^2} \right] t \right] \cr
&\times 
{e^{-2y/\ell_2} K_2(\ell_1|p|) K_2(\ell_2|p| e^{-y/\ell_2}) \over 
K_1(\ell_1|p|) K_2(\ell_2|p|) + K_1(\ell_2|p|) K_2(\ell_1|p|) + {1\over
2}b\ell |p|\, K_2(\ell_1|p|) K_2(\ell_2|p|)} \cr}\neqno\leftsolnR$$
for $y<0$.  

We first summarize the results that we derive below with a figure showing the
appropriate effective theory of gravity seen at different scales for different
relative choices for $\ell_1$, $\ell_2$ and $b\ell$.  For simplicity, the
lengths are assumed to be widely separated in scales in this figure so, for
example, in the top plot we assume $b\ell\ll\ell_1\ll\ell_2$.  The results
derived in the text are more general and often allow two of these lengths to
be of the same order.  Unless indicated otherwise, the effective theories
listed below are only valid at scales much larger or much smaller than the
bounding scale $\ell_1$, $\ell_2$ or $b\ell$.  
$$\beginpicture
\setcoordinatesystem units <1.0truein,1.0truein>
\setplotarea x from 0.0 to 4.0, y from -0.25 to 0.35
\put {length} [c] at 4.0 -0.2 
\put {$\ell_1$} [c] at 2.00 -0.15 
\put {$\ell_2$} [c] at 3.00 -0.15 
\put {$b\ell$} [c] at 1.00 -0.25 
\put {$4d$ gravity + } [c] at 0.5 0.325 
\put {scalar graviton} [c] at 0.5 0.175 
\put {$5d$ gravity} [c] at 1.5 0.25 
\put {$5d$ gravity} [c] at 2.5 0.25 
\put {$4d$ gravity} [c] at 3.5 0.25 
%
\setlinear
\arrow <7pt> [0.2,0.67] from 1.0 -0.2 to 1.0 0
\arrow <7pt> [0.2,0.67] from 3.75 0 to 4.0 0
\putrule from 0 0 to 4 0
\putrule from 2.00 0.05 to 2.00 -0.05
\putrule from 3.00 0.05 to 3.00 -0.05
\endpicture$$
$$\beginpicture
\setcoordinatesystem units <1.0truein,1.0truein>
\setplotarea x from 0.0 to 4.0, y from -0.25 to 0.35
\put {length} [c] at 4.0 -0.2 
\put {$\ell_1$} [c] at 1.00 -0.15 
\put {$\ell_2$} [c] at 3.00 -0.15 
\put {$b\ell$} [c] at 2.0 -0.25 
\put {$4d$ gravity + } [c] at 1.0 0.325 
\put {scalar graviton} [c] at 1.0 0.175 
\put {$5d$ gravity} [c] at 2.5 0.25 
\put {$4d$ gravity} [c] at 3.5 0.25 
%
\setlinear
\arrow <7pt> [0.2,0.67] from 2.0 -0.2 to 2.0 0
\arrow <7pt> [0.2,0.67] from 3.75 0 to 4.0 0
\arrow <7pt> [0.2,0.67] from 0.50 0.25 to 0.125 0.25
\arrow <7pt> [0.2,0.67] from 1.50 0.25 to 1.875 0.25
\putrule from 0 0 to 4 0
\putrule from 1.00 0.05 to 1.00 -0.05
\putrule from 3.00 0.05 to 3.00 -0.05
\endpicture$$
$$\beginpicture
\setcoordinatesystem units <1.0truein,1.0truein>
\setplotarea x from 0.0 to 4.0, y from -0.25 to 0.35
\put {length} [c] at 4.0 -0.2 
\put {$\ell_1$} [c] at 1.00 -0.15 
\put {$\ell_2$} [c] at 2.00 -0.15 
\put {$b\ell$} [c] at 3.00 -0.25 
\put {$4d$ gravity} [c] at 2.0 0.25 
\setlinear
\arrow <7pt> [0.2,0.67] from 3.0 -0.2 to 3.0 0
\arrow <7pt> [0.2,0.67] from 3.75 0 to 4.0 0
\arrow <7pt> [0.2,0.67] from 1.50 0.25 to 0.125 0.25
\arrow <7pt> [0.2,0.67] from 2.50 0.25 to 3.875 0.25
\putrule from 0 0 to 4 0
\putrule from 1.00 0.05 to 1.00 -0.05
\putrule from 2.00 0.05 to 2.00 -0.05
\endpicture$$
\smallskip

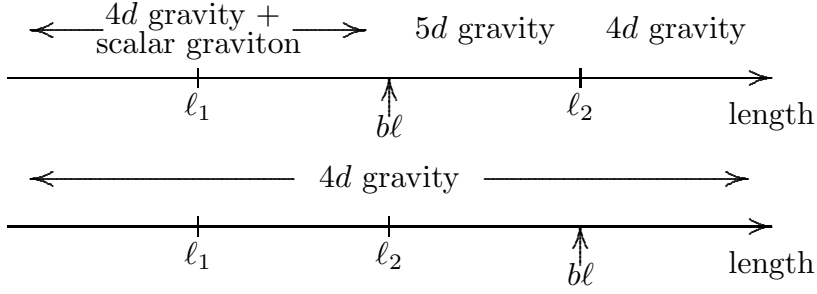
\centerline{ {\bf Figure 1.}  The effective classical theories of gravity for}
\centerline{different regions of the $(\ell_1, \ell_2, b\ell)$ parameter
space.}
\medskip

In the limit where both of the AdS lengths are small, $\ell_1|p|, \ell_2|p|
\ll 1$, the addition of a brane scalar curvature has the familiar $\ch$ effect
of rescaling the effective Newton's constant in the low energy theory.  To see
this effect, we note that when we substitute the leading behavior for
$\rightsolnR$,
$$\tilde\gamma_{\mu\nu}(p,0) = {16\pi\over p^2} \left[ t_{\mu\nu} - {1\over 3}
\left[ \eta_{\mu\nu} - {p_\mu p_\nu\over p^2} \right] t \right] 
{2\over \ell_1 + \ell_2 + b\ell} + \cdots , \neqno\zeromodeR$$
into the four dimensional Einstein tensor $\fourdEIN$ and use the trace in
$\gammatraceegR$, we obtain the correct form for four dimensional gravity,
$$\hat R_{\mu\nu} - {1\over 2}\eta_{\mu\nu} \hat R =
{16\pi\over\ell_1+\ell_2+b\ell} t_{\mu\nu} + \cdots . \neqno\fourdEINR$$
So for sufficiently large distances (the right side of Figure 1), we always
recover a standard theory of gravity.

The existence of a $4d$ curvature term in the action, however, allows us to
consider new corners of the $\ell_1$ and $\ell_2$ parameter space which did
not yield acceptable effective theories in the last section.  If we rewrite
the expression for $\tilde\gamma_{\mu\nu}$ on the brane as 
$$\tilde\gamma_{\mu\nu}(p,0) = {16\pi\over|p|} \left[ t_{\mu\nu} - {1\over 3}
\left[ \eta_{\mu\nu} - {p_\mu p_\nu\over p^2} \right] t \right] 
\left[ {1\over 2}b\ell |p| + {K_1(\ell_1|p|)\over K_2(\ell_1|p|)} +
{K_1(\ell_2|p|)\over K_2(\ell_2|p|)} \right]^{-1}  \neqno\bigblimit$$
we find that since $0\le K_1(z)/K_2(z)<1$ for $z\ge 0$, all that is needed to
recover a $1/r$ Newtonian potential is to have $b\ell|p|\gg 1$ so that 
$$\tilde\gamma_{\mu\nu}(p,0) = {16\pi\over p^2} \left[ t_{\mu\nu} - {1\over 3}
\left[ \eta_{\mu\nu} - {p_\mu p_\nu\over p^2} \right] t \right] {2\over b\ell}
+ \cdots \neqno\bigblimita$$
In this limit, both $\tilde\gamma_{\mu\nu}$ and $\gamma$ have a leading
$1/p^2$ behavior which automatically leads to a $1/r$ potential:
$$V(r) = - {1\over 3} {M\over M^3_5} {2\over b\ell} \left[ {4(\ell_1+\ell_2) +
3b\ell\over\ell_1 + \ell_2 + b\ell} \right] {1\over r} .
\neqno\potRbig$$
This equation is valid regardless of the scale of $\ell_1$ and $\ell_2$, 
compared with either the coefficient of the $4d$ curvature $(b\ell)$ or $r$. 
Yet when we use $\bigblimita$ with $\gammatraceegR$ to evaluate the Einstein
equation, we find that for arbitrary values of $\ell_1$ and $\ell_2$ that a
term involving the trace of the energy-momentum tensor for the brane fields
appears:
$$\hat R_{\mu\nu} - {1\over 2}\eta_{\mu\nu} \hat R = 
{16\pi\over b\ell} t_{\mu\nu} - {16\pi\over 3} {1\over b\ell} 
\left[ {\ell_1+\ell_2\over b\ell+\ell_1+\ell_2} \right] \left[ \eta_{\mu\nu} -
{\partial_\mu\partial_\nu\over\lform} \right]\, t + \cdots .
\neqno\fourdEINRbig$$
If $b\ell\gg\ell_1,\ell_2$, then the extra term can be neglected and the
leading behavior is that of a standard theory of four dimensional gravity,
$$\hat R_{\mu\nu} - {1\over 2}\eta_{\mu\nu} \hat R = 
{16\pi\over b\ell} t_{\mu\nu}  + \cdots . \neqno\fourdEINRgood$$
in agreement with $\fourdEINR$.  Note we have implicitly assumed that any
higher order terms in the effective theory of gravity on the brane $\baction$
can be neglected.

We can also study what happens when one or both of the AdS lengths becomes
infinite, while keeping $b\ell|p|\gg 1$.  These limits correspond respectively
to universes in which the brane is between a flat and an AdS region or is
simply embedded in a flat bulk space-time.  In either case, the effective
theory is 
$$\hat R_{\mu\nu} - {1\over 2}\eta_{\mu\nu} \hat R = 
{16\pi\over b\ell} \left[ t_{\mu\nu} - {1\over 3}\left( \eta_{\mu\nu} -
{\partial_\mu\partial_\nu\over\lform} \right)\, t \right] + \cdots .
\neqno\fourdEINRbad$$
The case of a brane embedded in a flat bulk was also investigated in $\dgp$
where the effective theory on the brane was shown to contain a scalar
graviton.  From $\fourdEINRbad$ it would appear that a scalar graviton is a
generic feature of branes embedded in (partially) flat space-times.

Thus far we have mapped the behavior of gravity when either
$\ell_1|p|,\ell_2|p|\ll 1$ or when $b\ell|p|\gg 1$ which covers most of figure
1.  What remains is the limit where $b\ell|p|\le 1$.  When $\ell_2|p|\gg 1
\gg\ell_1|p|$, then dominant contribution from $\bigblimit$ is 
$$\tilde\gamma_{\mu\nu}(p,0) = {16\pi\over|p|} \left[ t_{\mu\nu} - {1\over 3}
\left[ \eta_{\mu\nu} - {p_\mu p_\nu\over p^2} \right] t \right] 
 {1\over 1 + {1\over 2}(b\ell+\ell_1)|p| } + \cdots \neqno\smallblimitint$$
while for $\ell_2|p|,\ell_1|p|\gg 1$ we have
$$\tilde\gamma_{\mu\nu}(p,0) = {16\pi\over|p|} \left[ t_{\mu\nu} - {1\over 3}
\left[ \eta_{\mu\nu} - {p_\mu p_\nu\over p^2} \right] t \right] {1\over 2 +
{1\over 2} b\ell |p|} + \cdots . \neqno\smallblimitboth$$
In either case, we know from our experience in the previous section,
$\intmode$, that such a form for $\tilde\gamma_{\mu\nu}$ leads to a $1/r^2$,
or five dimensional, Newtonian potential.

\newsec{Conclusions.}

We have mapped out the behavior of the theory for different relative scales of
the AdS lengths in the bulk, $\ell_1$ and $\ell_2$, and the coefficient of the
brane curvature, $b\ell$, compared to experimentally testable lengths,
represented by $1/|p|$.  When both the AdS lengths are both smaller than a
millimeter, then regardless of the size of $b\ell$, the effective theory is
standard $4d$ gravity.  

With the scalar curvature term on the brane a new regime exists,
$\ell_1,\ell_2\ll b\ell$, in which a realistic classical theory of gravity
also emerges.  Interestingly we do not need to impose any further
constraints---provided $b\ell$ is sufficiently large, $\ell_1$ and $\ell_2$ do
not need to be below the millimeter limit.  At distances of the order of
$\ell_1,\ell_2$ or less, the brane curvature term dominates while at large
distances, both the brane curvature and the effective bulk theory contribute
to a standard $4d$ theory of gravity.  The largest phenomenologically
acceptable AdS$_5$ lengths depend in turn upon the largest plausible value for
$b\ell$.  In terms of $M_4$, the effective $4d$ Planck mass, we can estimate
$b\ell$ using $\fourdEINRgood$:
$$b\ell = {2M_4^2\over M_5^3}. \neqno\plancks$$
For example, when $M_5=1\, {\rm TeV}$, then $b\ell$ is about 400 AU.  Thus the
theory can accommodate AdS lengths much larger than the current millimeter
bound and still lie safely in the $\ell_1,\ell_2\ll b\ell$ regime. 

In this regime, if $b$ is not unnaturally large, then a clear experimental
signature for this universe would be the appearance of a weak correction to
the Einstein equation that depends on the trace of the energy-momentum tensor
$\fourdEINRbig$.  In terms of the parameterized post-Newtonian formalism
$\mtw$, this extra term in the effective field equations leads to a
perturbation in the parameter $\gamma$ which measures how the curvature of
space depends on the presence of a rest mass.  To leading order, 
$$\gamma - 1 = - {2\over 3} {\ell_1+\ell_2\over b\ell} + \cdots .
\neqno\cfppn$$
The current value of $|\gamma-1|$ based on VLBI observations is less than $3
\times 10^{-4}$ 
\ref\will{C.~M.~Will, {\it Theory and Experiment in Gravitational Physics\/,}
Cambridge University Press:  Cambridge (1993)
and 
C.~M.~Will, ``The confrontation between general relativity and experiment:  A
1998  update,'' gr-qc/9811036.}.  
Thus for a symmetric universe in which $\ell_1=\ell_2=\ell$, this limit
constrains $b\ge 4000$.  Alternatively if $M_5\approx 1$ TeV, then from
$\plancks$ the model can accommodate AdS lengths as large as $10^{10}$ m.

We have here extended the approach of using a gauge in which the position of
the brane remains fixed $\gr$ to allow an unambiguous statement of the
boundary conditions in a setting more general than the standard orbifold
universes.  A correct understanding of these boundary conditions is crucial
for determining the form of the effective theory of gravity on the brane.  Our
results can be readily generalized.  For example, the brane itself might be
given some non-trivial global curvature as in 
\ref\per{P.~Kraus, ``Dynamics of anti-de Sitter domain walls,'' JHEP {\bf
9912}, 011 (1999) [hep-th/9910149].}  
and $\ch$ where the AdS bulk metric is replaced with an AdS-Schwarzschild
metric.  It is also straightforward to include more general actions on the
brane; their existence would simply add extra terms to the right side of the
Israel equation, $\israelall$.

\appendix{A}{Conventions and Components.}

The conventions that we have used throughout use a metric with signature
$(-,+,\cdots,+)$ and a Riemann curvature tensor defined by 
$$-R^a_{\ bcd} \equiv \partial_d \Gamma^a_{bc} - \partial_c \Gamma^a_{bd} +
\Gamma^a_{ed}\Gamma^e_{bc} - \Gamma^a_{ec}\Gamma^e_{bd} . \neqno\Riemann$$
${\cal R}^a_{\ bcd}$ and $\hat R^\rho_{\ \mu\nu\lambda}$, the curvature
tensors corresponding respectively to the induced metric $h_{ab}$ or its
transverse components $\hat g_{\mu\nu}$, use the same sign convention.

In extracting the terms in the Einstein equation and the Israel condition,
which are linear in the perturbation about the background metric, we have used
the following expressions for the Christoffel symbols for the induced metric, 
$$\hat\Gamma^\lambda_{\mu\nu} = {1\over 2} \left[ \partial_\mu
\gamma^\lambda_\nu + \partial_\nu \gamma^\lambda_\nu - \partial^\lambda
\gamma_{\mu\nu} \right] . \neqno\chirsind$$
while for the bulk metric we have 
$$\eqalign{
\Gamma^\lambda_{\mu\nu} &= \hat\Gamma^\lambda_{\mu\nu} + {1\over 2}{f'\over f}
n^\lambda\eta_{\mu\nu} \cr
\Gamma^y_{\mu\nu} &= - {1\over 2} {1\over n^2} f' \eta_{\mu\nu} - {1\over 2}
\left[ f'\gamma_{\mu\nu} + f \partial_y \gamma_{\mu\nu} \right] + {1\over 2}
\left( \partial_\mu n_\nu + \partial_\nu n_\mu \right) \cr
\Gamma^\lambda_{\mu y} &= {1\over 2} {f'\over f} \eta^\lambda_\mu + {1\over 2}
\partial_y \gamma^\lambda_\mu + {1\over 2}{1\over f} \left( \partial_\mu
n^\lambda - \partial^\lambda n_\mu \right) \cr
\Gamma^y_{\mu y} &= - {1\over 2} {f'\over f} n_\mu + {1\over 2}
\partial_\mu\phi \cr
\Gamma^\lambda_{yy} &= {1\over 2}{1\over f} \left[ 2 \partial_y n^\lambda -
\partial^\lambda\phi \right] \cr
\Gamma^y_{yy} &= {1\over 2} \partial_y\phi .\cr }\neqno\chris$$

The linear contributions to the transverse components of the Ricci tensor for
the bulk and the induced metrics are respectively
$$\eqalign{
R_{\mu\nu} 
&= \hat R_{\mu\nu} + {1\over 2} \left[ f^{\prime\prime} + {f^{\prime 2}\over
f} \right] \left[ {1\over n^2} \eta_{\mu\nu} + \gamma_{\mu\nu} \right] +
{1\over 2} \partial_\mu\partial_\nu\phi + {1\over 2} f
\partial_y^2\gamma_{\mu\nu} \cr
&\quad 
+ {1\over 2} f' \left[ 2 \partial_y \gamma_{\mu\nu} + {1\over 2} \eta_{\mu\nu}
\partial_y \gamma - {1\over 2} \eta_{\mu\nu} \partial_y\phi \right] \cr
&\quad
- {1\over 2} {f'\over f} \left[ \partial_\mu n_\nu + \partial_\nu n_\mu +
\eta_{\mu\nu}\partial^\lambda n_\lambda \right] - {1\over 2} \partial_y \left(
\partial_\mu n_\nu + \partial_\nu n_\mu \right) \cr
\hat R_{\mu\nu} &= {1\over 2} \left[ \partial_\mu\partial_\nu\gamma +
\partial^\lambda\partial_\lambda\gamma_{\mu\nu} - \partial_\mu
\partial^\lambda \gamma_{\lambda\nu} - \partial_\nu \partial^\lambda
\gamma_{\mu\lambda} \right] .\cr} \neqno\curves$$
Finally, the linear terms in the extrinsic curvature are 
$$K^\mu_\nu = {1\over 2n} \left[ {f'\over f} \eta^\mu_\nu +
\partial_y\gamma^\mu_\nu - {1\over f} \left( \partial^\mu n_\nu + \partial_\nu
n^\mu \right) \right] \neqno\extfirst$$
where $f(y)$ is the appropriate warp factor for either bulk region.

\listrefs
\bye